\documentclass[
twocolumn
]{aastex631}
\usepackage{graphics} 
\usepackage{epsfig} 
\usepackage{mathptmx} 
\usepackage{times} 
\usepackage{amsmath} 
\usepackage{amssymb}  
\usepackage{xcolor}
\usepackage{subfigure}
\usepackage{hyperref}

\usepackage{bm}
\graphicspath{{./}}

\submitjournal{ApJL}

\shorttitle{Isotropization and Evolution of Energy-Containing Eddies}
\shortauthors{}

\begin{document}

\title{Isotropization and Evolution of Energy-Containing Eddies in Solar Wind Turbulence: Parker Solar Probe, Helios 1, ACE, WIND, and Voyager 1}

\correspondingauthor{Manuel Enrique Cuesta}
\email{mecuesta@udel.edu}

\author[0000-0002-7341-2992]{Manuel Enrique Cuesta}
\affiliation{University of Delaware,
Department of Physics and Astronomy,
Newark, DE 19716, USA}

\author[0000-0002-7174-6948]{Rohit Chhiber}
\affiliation{University of Delaware,
Department of Physics and Astronomy,
Newark, DE 19716, USA}
\affiliation{NASA Goddard Space Flight Center,
Greenbelt, MD 20771, USA}

\author[0000-0003-3891-5495]{Sohom Roy}
\affiliation{University of Delaware,
Department of Physics and Astronomy,
Newark, DE 19716, USA}

\author{Joshua Goodwill}
\affiliation{University of Delaware,
Department of Physics and Astronomy,
Newark, DE 19716, USA}

\author[0000-0003-4168-590X]{Francesco Pecora}
\affiliation{University of Delaware,
Department of Physics and Astronomy,
Newark, DE 19716, USA}

\author{Jake Jarosik}
\affiliation{University of Delaware,
Department of Physics and Astronomy,
Newark, DE 19716, USA}

\author[0000-0001-7224-6024]{William H. Matthaeus}
\affiliation{University of Delaware,
Department of Physics and Astronomy, Bartol Research Institute, Newark, DE 19716, USA}

\author[0000-0003-0602-8381]{Tulasi N. Parashar}
\affiliation{Victoria University of Wellington,
Wellington, New Zealand, 6012}

\author[0000-0002-6962-0959]{Riddhi Bandyopadhyay}
\affiliation{Princeton University,
Department of Astrophysical Sciences,
Princeton, NJ 08544, USA}



\begin{abstract}

We examine the radial evolution of correlation lengths perpendicular (\(\lambda_C^{\perp}\)) and parallel (\(\lambda_C^{\parallel}\)) to the magnetic-field direction, computed from solar wind magnetic-field data measured by Parker Solar Probe (PSP) during its first eight orbits, Helios 1, Advanced Composition Explorer (ACE), WIND, and Voyager 1 spacecraft.
Correlation lengths are grouped by an interval's alignment angle; the angle between the magnetic-field and solar wind velocity vectors (\(\Theta_{\rm BV}\)). 
Parallel and perpendicular angular channels correspond to angles \(0^{\circ}~<~\Theta_{\rm BV}~<~40^{\circ}\) and \(50^{\circ}~<~\Theta_{\rm BV}~<~90^{\circ}\), respectively.
We observe an anisotropy in the inner heliosphere within 0.40~au, with  \(\lambda_C^{\parallel} / \lambda_C^{\perp} \approx 0.75\) at 0.10~au. 
This anisotropy reduces with increasing heliocentric distance and the correlation lengths roughly isotropize within 1~au.
Results from ACE and WIND support a reversal of the anisotropy, such that \(\lambda_C^{\parallel} /\lambda_C^{\perp} \approx 1.29\) at 1~au. 
The ratio does not appear to change significantly beyond 1~au, although the small number of parallel intervals in the Voyager dataset precludes unambiguous conclusions from being drawn. 
This study provides insights regarding the radial evolution of the large, most energetic interacting turbulent fluctuations in the heliosphere. 
We also emphasize the importance of tracking the changes in sampling direction in PSP measurements as the spacecraft approaches the Sun, when using these data to study the radial evolution of turbulence. 
This can prove to be vital in understanding the more complex dynamics of the solar wind in the inner heliosphere and can assist in improving related simulations.

\end{abstract}



\section{Introduction} \label{sec:intro}

    Plasma turbulence in the magnetohydrodynamic (MHD) regime has a well-known tendency to develop and sustain anisotropy relative to the mean magnetic-field direction \citep[e.g.,][]{oughton2015RSPTA}.
    This anisotropy has been extensively studied in observational, experimental, theoretical, and numerical works \citep[e.g.,][]{robinson1971PhFl,ShebalinEA1983,MatthaeusEA1990JGR,goldreich1995ApJ,DassoEA2005,chhiber2020PoP}, and has significance for heliospheric plasma dynamics \citep{deForestEA2016}, turbulence transport \citep{ZankEA2021PoP}, and energetic particle scattering \citep{OughtonEA2021}. Studies of anisotropy in the solar wind have most often concentrated on {\it spectral} (i.e.,     correlation) anisotropy or {\it polarization} ({\it variance}) anisotropy, in each case as measured in the inertial range \citep{oughton2015RSPTA}. Generally speaking, the larger, outer scale, or energy containing eddies are expected to exhibit less anisotropy
    \citep{goldreich1995ApJ}. Nevertheless it is of interest to examine the dynamical development of outer-scale anisotropy, especially in Parker Solar Probe (PSP) spacecraft data \citep{FoxEA2016}, which may provide valuable insights concerning the dynamics of the young solar wind. 
    Here we examine the radial evolution  of large-scale anisotropies -- at the scale of the correlation length -- as observed by PSP, complemented by other spacecraft at larger distances.

    With each additional orbit, PSP compiles 
    measurements of solar wind plasma in previously unexplored regions.
    In the inner heliosphere, the distinctive features of PSP's orbit implies 
    sampling directions along the direction of bulk plasma flow in the spacecraft frame that differ from 
    earlier spacecraft. 
    The directions of the flow and the magnetic field as viewed by the spacecraft can be important when determining whether observed fluctuations of measured quantities are varying either parallel or perpendicular to the magnetic field.
    The angle between the flow and magnetic-field vectors (\textit{alignment angle}, \(\Theta_{\rm BV}\)) varies
    mainly due to the change in the heliospheric magnetic-field direction between PSP aphelia and perihelia, as well as the changes in the spacecraft velocity throughout the orbit.
    At greater distance from the sun, the spacecraft speed is smaller, and the magnetic field direction, while still varying, is much less radial than at PSP perihelia.
    
    The Parker spiral average magnetic-field \citep{Parker1958} organizes the baseline 
    trend of these angles with varying radial distance.
    However, for PSP's closest approaches to the sun, both the 
    flow and the magnetic field are dominantly radial and
    PSP most often measures variations parallel to the magnetic field, yielding a deficit in measurements 
    perpendicular to the magnetic field.
    This calls for care in interpretation of PSP observations, since 
    the observed correlations may not be representative of the entire system.
    Here we examine separately the radial evolution of parallel and perpendicular 
    energy-containing correlation scales. 
    Parallel and perpendicular angular channels are chosen to correspond to angles \(0^{\circ}~<~\Theta_{\rm BV}~<~40^{\circ}\) and \(50^{\circ}~<~\Theta_{\rm BV}~<~90^{\circ}\), respectively.
    
    To expand the scope of the study, we employ observations by PSP, Helios 1, Advanced Composition Explorer (ACE), WIND, and Voyager 1.
    Previous studies have investigated the relationship between the parallel and perpendicular correlation scales (\(\lambda_C^{\parallel}\) and \(\lambda_C^{\perp}\)).
    \citet{RuizEA2011} observed an anisotropy using Helios 1 data such that \(\lambda_C^{\parallel}<\lambda_C^{\perp}\), whereas others \citep{MatthaeusEA1990JGR,DassoEA2005,WeygandEA2011} reported that parallel lengths are greater near \(1~{\rm au}\).
    More recently, PSP observes in its first five orbits that \(\lambda_C^{\parallel}<\lambda_C^{\perp}\) for heliocentric distances \(R<0.30~{\rm au}\) \citep{BandyopadhyayMcComas2021}.
    
    We are unable to separately study fast and slow wind intervals here, with the exception of ACE and WIND datasets.
    The other datasets used in this study are dominated by slow wind (\(V_{\rm SW}<450~{\rm km/s}\)), yielding weak statistical significance for results in fast wind (\(V_{\rm SW}>600~{\rm km/s}\)).
    With this limitation in mind, we find evidence for the isotropization of \(\lambda_C^{\parallel}\) and \(\lambda_C^{\perp}\) with increasing \(R\).
    Once isotropy is achieved, it continues to evolve depending on system dynamics causing temporary deviations; however, we cannot confidently comment on this evolution beyond \(1~{\rm au}\) due to weak statistical significance of parallel intervals observed by Voyager 1.
    In Section \ref{sec:methods}, we explain the methods used to compute the correlation length.
    Afterwards, we present the radial evolution of the winding angle and correlation lengths in Sections \ref{sec:angles} and \ref{sec:corr_lengths}, respectively, with further discussion in Section \ref{sec:disc}.
    We include instrumentation and data specifics in Appendix \ref{sec:data}.

\section{Analytic Methods}\label{sec:methods}

    We define magnetic field fluctuations \(\bm{b}\) by subtracting the averaged magnetic field from the total magnetic field \(\bm{B} - \langle \bm{B} \rangle\), where \(\langle\cdot\rangle\) refers to a temporal average over an appropriately sized interval.
    For interval duration information, see Appendix \ref{sec:data}.
    The Taylor ``frozen-in'' hypothesis (TH) \citep{Taylor1938} implicitly associates a temporal lag \(\tau\) with a spatial lag \(\ell\) according to \(\ell = V_{\rm SW}\tau\),  where \(V_{\rm SW}\) is the bulk flow speed in the spacecraft frame, computed over a specified averaging interval.
    This approximate conversion from temporal to spatial lags is expected to be accurate when \(V_{\rm SW}\) is large compared to characteristic speeds of the local fluctuation dynamics, such as the rms fluctuation speed or Alfv\'en speed \(V_{\rm A} = B/ \sqrt{4\pi \rho_i}\), where \(B\) is the magnetic-field vector magnitude and \(\rho_i\) is the ion mass density.
    At distances near PSP perihelia, \(V_{\rm SW}\) is comparable to \(V_{\rm A}\), weakening the validity of TH \citep{ChhiberEA2019May,PerezEA21}.
    In this study, \(\sim 1\%\) of PSP intervals (see Appendix \ref{sec:data}), encompassing only radial distances \(R<0.20~{\rm au}\), exhibit \(V_{\rm A}/V_{\rm SW} > 0.66\), which we considered to be poor validity of TH.
    Only \(\sim 0.10\%\) of intervals have \(V_{\rm A}/V_{\rm SW} > 1\). Therefore, TH remains at either intermediate or high validity for the large majority of this study.

    \textit{Autocorrelation.}
    The two-time autocorrelation function \(R_C(\tau)\) of time-stationary magnetic-field fluctuations is defined as:
    \begin{equation}\label{eq:corr}
        R_C(\tau) = \frac{\langle \bm{b}(t) {\bf \cdot} \bm{b}(t+\tau) \rangle}{\langle \bm{b}(t) {\bf \cdot} \bm{b}(t) \rangle}.
    \end{equation}
    %
    The correlation time \(\tau_e\) is the characteristic time separation over which the fluctuations become uncorrelated. 
    Here we identify \(\tau_e\) with the ``e-folding'' time, i.e., \(R_C(\tau = \tau_e) = 1/e\).
    This correlation time corresponds to the size of the energy-containing eddies, within the interval used for computation.
    Here TH is used to convert \(\tau_e\) to a spatial correlation scale, such that \(\lambda_C = V_{\rm SW}\tau_e\).
    
    For all spacecraft in this study, with the exception of ACE, we use this method to estimate
    \(\lambda_C\).
    For ACE, we employ \(\lambda_C\) values 
    obtained from \citep{RoyEA2021}, and computed via exponential fit
    of the correlation function. 
        A preliminary estimate \(\tau_e^\prime\) of the correlation time is first obtained via the ``e-folding'' method, given by \(R(\tau_e^{\prime})/R(0)=1/e\).
    Then a linear least-squares fit to \(\log \left[ R(\tau)/R(0) \right] \sim -{\tau }/{\tau_e}\) is performed over the interval \(\tau \in \left[0,\tau_e^{\prime}/2 \right]\) to compute \(\tau_e\), which can is converted to a 
    corresponding spatial scale \(\lambda_C\), as above.

    \begin{figure*}[htp!]
        \centering
        \includegraphics[scale=.75]{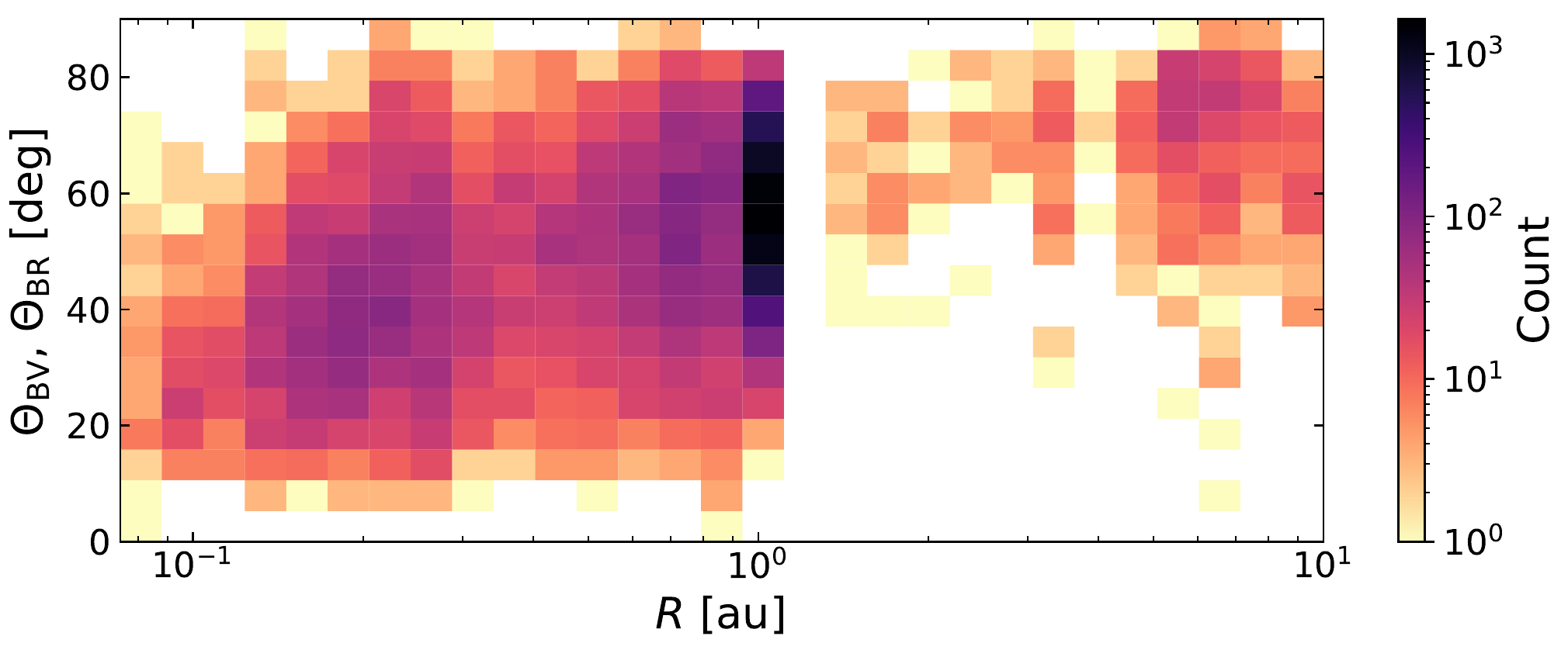}
        \caption{Radial distributions of alignment angle \(\Theta_{\rm BV}\) for PSP and \(\Theta_{\rm BR}\) for ACE, WIND, and Voyager 1. PSP interval sizes are 1~hour and 3~hours for heliocentric distances \(R<0.3~{\rm au}\) and \(0.3~{\rm au}<R<1~{\rm au}\), respectively.  Single day intervals are used for ACE, WIND, and Voyager 1 for \(R\ge 1~{\rm au}\). Interval count of each bin is keyed to color bar. White bins denote zero count.}
        \label{fig:theta_r}
    \end{figure*}


\section{Radial Distribution of Alignment Angle}\label{sec:angles}

    A key parameter in this study is the angle between the magnetic-field and flow velocity vectors, denoted \(\Theta_{\rm BV}\).
    In most cases, the radial velocity component dominates the tangential and normal components, thus motivating an approximation, and simplification, of this alignment angle when appropriate.
    This approximation leads to the \textit{winding angle} \(\Theta_{\rm BR}\) representing the angle between the magnetic-field and radial unit vectors defined as
    
    \begin{equation}
        \Theta_{\rm BR} = \cos^{-1} \left( \frac{\langle|B_R|\rangle}{\langle\parallel \bm{B} \parallel\rangle} \right)
    \end{equation}
    where \(B_R\) is the radial component of the magnetic field \(\bm{B}\) in a heliocentric RTN coordinate system \citep{Franz2002PSS},
    \(| \cdot |\) is an absolute value, and \(\parallel \cdot \parallel\) is a vector magnitude.
    The winding angle may also be referred to as the alignment angle for convenience.
    Taking the absolute value of \(B_R\) is necessary to avoid its average from vanishing in intervals that include crossings of the heliospheric current sheet (HCS) with an associated polarity reversal. If this operation is not performed prior to averaging then an interval with a HCS crossing may be improperly labeled as a perpendicular interval.
    
    For ACE, WIND, and Voyager 1, it is sufficient to examine \(\Theta_{\rm BR}\) since the observed direction of the flow is dominantly radial and spacecraft speeds are negligible compared to the bulk flow speed.
    However in PSP data the spacecraft velocity and the tangential component of the solar wind velocity can be comparable to the radial component of the latter \citep{FoxEA2016,Kasper2019Nature}, so the angle between the magnetic field and the sampling direction is no longer well-represented by \(\Theta_{\rm BR}\). 
    Therefore, we compute \(\Theta_{\rm BV}\) directly, defined as 
    
    \begin{equation}
        \Theta_{\rm BV} = \cos^{-1} \left( \frac{ \langle|\bm{B}|\rangle \cdot \langle\bm{V}_{\rm SW}\rangle}{\langle \parallel \bm{B} \parallel\rangle \langle\parallel \bm{V}_{\rm SW} \parallel\rangle} \right),
    \end{equation}
    where \(V_{\rm SW}\) is the solar wind velocity measured in the spacecraft frame.
    For any intervals in which the computation of \(\Theta_{\rm BV}\) is not possible (due to missing velocity data), then \(\Theta_{\rm BR}\) will be used instead.
    Additionally, we constrain the alignment angles to lie in the range between \(0^{\circ}\) and \(90^{\circ}\) by not distinguishing between parallel or anti-parallel. 
    
    The radial distributions of alignment angles are presented in Figure \ref{fig:theta_r} for all spacecraft excluding Helios 1.\footnote{\citet{RuizEA2011} show the angles between the magnetic-field and flow vectors for Helios 1; however, their parallel and perpendicular angular channels are narrower than those used in the present study.}
    The count in each bin is keyed to the color bar.
    We observe a radially increasing alignment angle, corresponding to 
    increasing central density in vertical slices.
    This is consistent with the mean Parker spiral magnetic field.  
    We have sufficient coverage of both angular channels, \(0^{\circ}~<~\Theta_{\rm BV}~<~40^{\circ}\) and \(50^{\circ}~<~\Theta_{\rm BV}~<~90^{\circ}\), for \(R<1~{\rm au}\).
    However, Voyager 1 has a narrower range of angular coverage, since the mean Parker-spiral magnetic field forms an increasingly large angle relative to the radial (flow) direction, at those distances.

    \begin{figure*}[htp!]
        \centering
        \includegraphics[scale=.75]{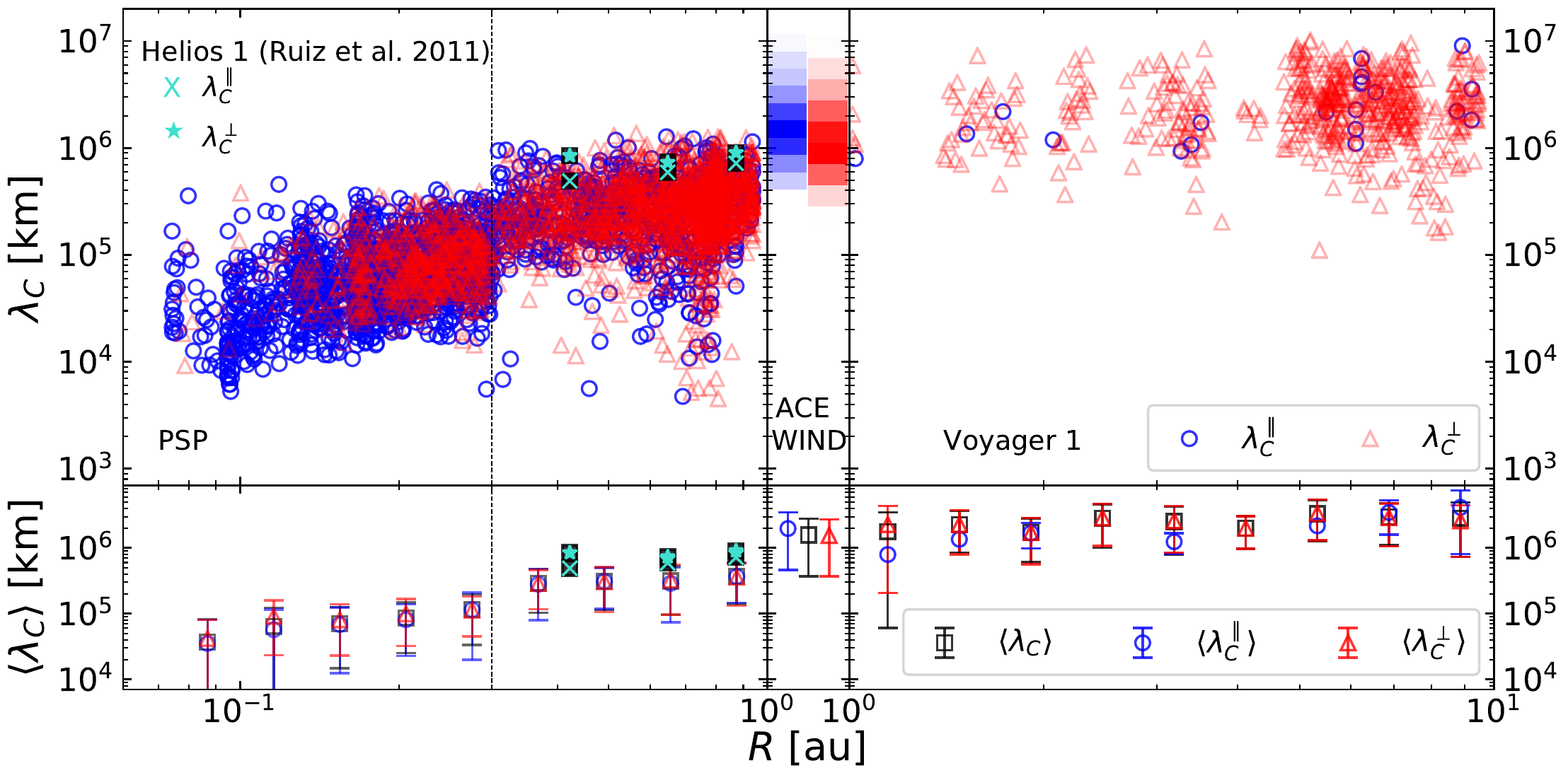}
        \caption{Correlation lengths observed by PSP and Helios 1 (left), ACE and WIND (middle), and Voyager 1 (right), 
        vs. heliocentric distance. (Top)  results for individual intervals. (Bottom) binned averages with bars representing standard deviation about mean. Blue circles, red triangles, and black squares represent parallel, perpendicular, and all intervals, respectively. Helios 1 data represented by three turquoise crosses (\(\lambda_C^{\parallel}\)) and stars (\(\lambda_C^{\perp}\)) are extracted from \citet{RuizEA2011}. PSP data to 
        left and right of vertical dashed line at 0.3 au represent 1-hour and 3-hour interval sizes, respectively. 
        Columns at \(\sim 1~{\rm au}\) represent number density (darker shades = larger counts) 
        of 1-day intervals for ACE and WIND grouped as \(\lambda_C^{\parallel}\) (left, blue) and \(\lambda_C^{\perp}\) (right, red).
        Statistics of these distributions are in Table \ref{tab:acewind4050}.
        Power-law fits for \(\lambda_C\), \(\lambda_C^{\parallel}\), and \(\lambda_C^{\perp}\) are in Table \ref{tab:rad_scaling}.}
        \label{fig:corr_wide}
    \end{figure*}
    
\section{Radial Variation of Parallel and Perpendicular Correlation Lengths}\label{sec:corr_lengths}
    
    Using results from PSP, Helios 1, ACE, WIND, and Voyager 1, we compare the evolution of perpendicular and parallel correlation lengths from \(\sim 0.08~{\rm au}\) (16~\(R_\odot\)) to \(10~{\rm au}\).
    We illustrate the results in Figure \ref{fig:corr_wide}.
    The top row of panels shows the radial variation of \(\lambda_C^{\parallel}\) and \(\lambda_C^{\perp}\); 
    the bottom row of panels shows averages of these quantities within radial bins equally-spaced in  log \(R\). The density of blue and red points (top) 
    demonstrates the transition from dominant-parallel sampling close to the Sun to dominant-perpendicular sampling above 1~au, as also seen in Figure \ref{fig:theta_r}. Both correlation scales systematically increase with \(R\) 
    by nearly two orders of magnitude from \(\sim 5\times10^4\) km at 0.10~au to \(\sim 3\times 10^6\) km at 10~au. The values between 0.40~au to 5~au are consistent with previous work using Helios, ACE, and Ulysses observations \citep{Ruiz2014SP}. 
    The increase in \(\lambda_C\) reflects the ``aging'' of turbulence, with larger scales participating in the turbulent cascade as the solar wind evolves  \citep{Matthaeus1998JGR_turbulence_age,Bruno2013LRSP} and 
    flux tubes expand \citep{Hollweg1986JGR}. 
    Radial power-law fits are presented in Table \ref{tab:rad_scaling}. 
    The discontinuity at \(R=0.30~{\rm au}\) is 
    due to a shift from a 1-hr to 3-hr interval duration. 
    The effect of interval size on the correlation scale is well-known \citep{IsaacsEA2015} and can also be seen at the 1 au boundary between PSP and ACE/WIND data, where we change from 3-hr to 1-day intervals.
    
\begin{figure}[htp!]
    \centering
    \includegraphics[scale=.7]{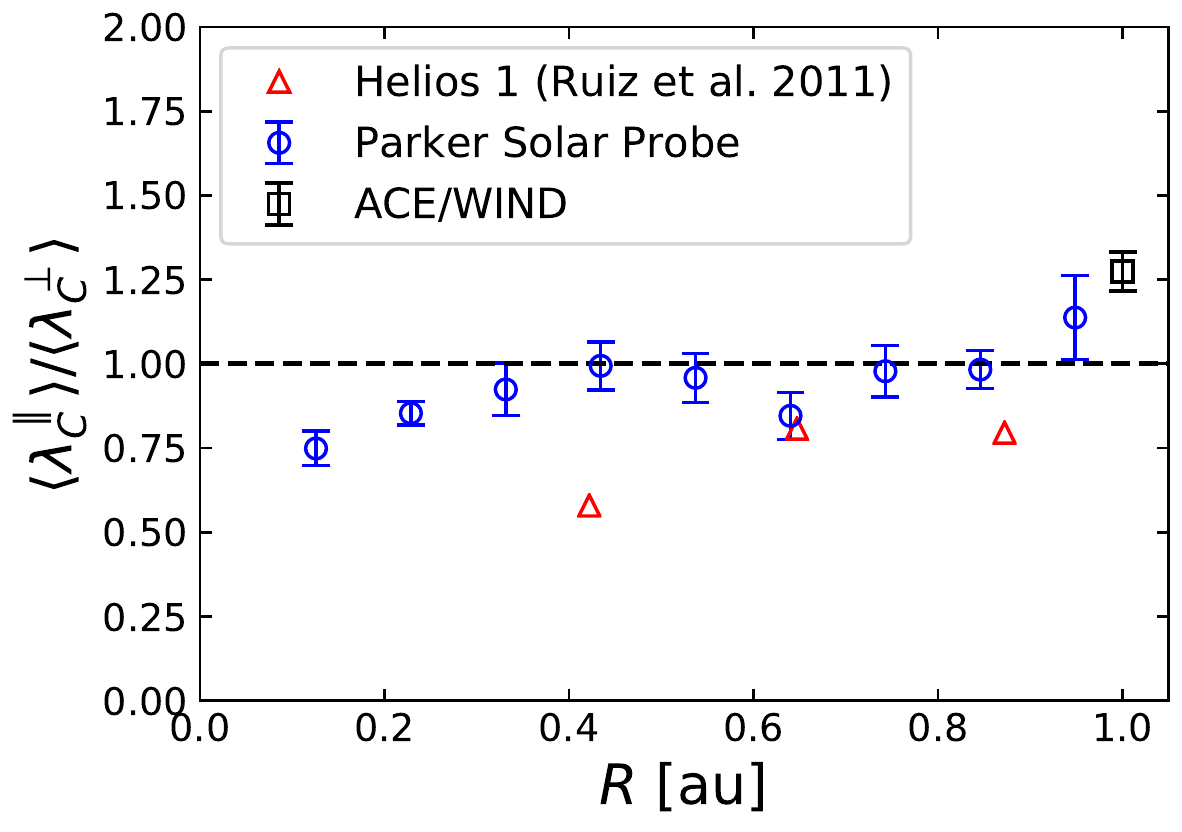}
    \caption{Ratios of \(\langle \lambda_C^{\parallel} \rangle\)/\(\langle \lambda_C^{\perp} \rangle\) for PSP (blue circles), Helios 1 (red triangles), and ACE/WIND (black square). Points represent radially binned averages of results presented in Figure \ref{fig:corr_wide}. The statistics of the distribution from ACE/WIND are given in Table \ref{tab:acewind4050}. Error-bars represent standard error of the mean, \(\sigma/\sqrt{n}\), where \(\sigma\) is standard deviation and \(n\) is number of samples. We also compute an average ratio \(\langle \lambda_C^{\parallel} \rangle\)/\(\langle \lambda_C^{\perp} \rangle \approx 0.97\)
    ~ over all Voyager data out to 10 au 
    , with 0.17 
    standard error. }
    \label{fig:ratio_wide}
    \end{figure}

    For a quantitative examination of the radial evolution of anisotropy, we compute \(\langle \lambda_C^{\parallel} \rangle / \langle \lambda_C^{\perp} \rangle\), where each of the correlation scales are first radially averaged in bins of size \(0.10~{\rm au}\), for \(R \leq 1~{\rm au}\).
    These accumulated averages are shown in Figure \ref{fig:ratio_wide}, which demonstrates the radial evolution of the observed anisotropy in the inner heliosphere.
    Data above 1~au are not shown because parallel intervals have low statistical weight.    
    Nevertheless, for the full collection of samples 
    beyond 1~au, we compute an overall average value 
    \(\langle \lambda_C^{\parallel} \rangle\)/\(\langle \lambda_C^{\perp} \rangle~\approx~0.97\pm 0.17\)
     ~ for the heliocentric distances covered by Voyager 1. 
    Therefore, we observe a continued evolution 
    in the outer heliosphere that may be characterized as either mild anisotropy or approximate  isotropy.
    
    Figure \ref{fig:ratio_wide} shows that a trend toward increasing anisotropy develops below \(\approx 0.4~{\rm au}\), with  \(\lambda_C^{\parallel} < \lambda_C^{\perp}\).
    This result, also seen by \cite{Bandyopadhyay2021ApJ_geometry}, appears superficially to be in contrast 
    to interpretations of images obtained by STEREO \citep{deForestEA2016} 
    that demonstrate a transition from {\it striated} thread-like morphology to more 
    isotropic {\it flocculated} patterns. 
    It has been argued that this transition from coronal quasi-two dimensional structure \citep{ZankEA2021PoP}
    to more isotropic turbulence outside the Alfv\'en critical zone is driven by 
    dynamics of microstream shears \citep{RuffoloEA2020}.  
    As noted by \citet{deForestEA2016}, fluctuations at the 
    correlation scale are much smaller than structure detected in the imaging studies.
    Therefore there is no direct contradiction, though the origins of initially very small parallel correlation scales remains unexplained. 
    The picture that emerges is of a faster increase of \(\lambda_C^{\parallel}\) up to \(\approx 0.40~{\rm au}\), which can be observed in Figure \ref{fig:ratio_wide}.
    Possible reasons for this more rapid increase in \(\lambda_C^{\parallel}\) compared to \(\lambda_C^{\perp}\) are discussed in greater detail in Section \ref{sec:disc}.
    
    \begin{table*}[htp!]
    \centering
    \begin{tabular}{c|c|c|c}
         & \(\langle\lambda_C^{\parallel}\rangle\) [\(10^6~{\rm km}\)] & \(\langle\lambda_C^{\perp}\rangle\) [\(10^6~{\rm km}\)] & \(\langle\lambda_C^{\parallel}\rangle/\langle\lambda_C^{\perp}\rangle\) \\
         \hline
        All \(V_{\rm SW}\)            & \(1.98 \pm 0.09\) & \(1.55 \pm 0.02\) & \(1.28 \pm 0.06\) \\
        \(V_{\rm SW}<450~{\rm km/s}\) & \(1.98 \pm 0.11\) & \(1.67 \pm 0.02\) & \(1.19 \pm 0.07\) \\
        \(V_{\rm SW}>600~{\rm km/s}\) & \(1.78 \pm 0.31\) & \(1.05 \pm 0.05\) & \(1.70 \pm 0.31\)
    \end{tabular}
    \caption{Combined ACE and WIND statistics 
     for \(\lambda_C^{\parallel}\) and \(\lambda_C^{\perp}\), differentiated by wind speed. 
     Values represent the mean value along with standard error of the mean, \(\sigma/\sqrt{n}\), where \(\sigma\) is standard deviation and \(n\) is number of samples.}
    \label{tab:acewind4050}
    \end{table*}
    
    \begin{table*}[htp!]
    \centering
    \begin{tabular}{c|c|c|c}
        \(\lambda_C \sim R^{\alpha \pm \sigma}\) & \(R<0.30~{\rm au}\) & \(0.30~{\rm au}<R<1.0~{\rm au}\) & \(R>1~{\rm au}\) \\
         \hline
        \(\lambda_C\)             & \(0.97 \pm 0.04\) & \(0.29 \pm 0.01\) & \(0.27 \pm 0.01\) \\
        \(\lambda_C^{\parallel}\) & \(1.03 \pm 0.06\) & \(0.28 \pm 0.03\) & \(0.64 \pm 0.12\) \\
        \(\lambda_C^{\perp}\)     & \(0.61 \pm 0.06\) & \(0.27 \pm 0.02\) & \(0.23 \pm 0.01\)
    \end{tabular}
    \caption{Radial power-law fits to \(\lambda_C\), \(\lambda_C^{\parallel}\), and \(\lambda_C^{\perp}\) from Figure \ref{fig:corr_wide}. Quantities represent fitted parameters \(\alpha\) with standard deviation \(\sigma\) about the best fit.}
    \label{tab:rad_scaling}
    \end{table*}
    
    After attaining isotropy near 0.40~au, the turbulence  continues to evolve with increasing radius, with some variation towards lower \(\lambda_C^{\parallel}\) near 0.60~au. 
    The ratio then attains isotropy again by 0.8~au and increases such that \(\lambda_C^{\parallel} > \lambda_C^{\perp}\) near 1~au. 
    This situation persists at 1~au where similar results are observed in ACE and WIND data.  
    Table \ref{tab:acewind4050} shows averages and standard deviations of the ACE and WIND observations that are shown as distributions in Figure \ref{fig:corr_wide}.
    \citet{DassoEA2005} report that \(\langle \lambda_C^{\parallel} \rangle/\langle \lambda_C^{\perp} \rangle = 0.71\) for \(V_{\rm SW} > 600~{\rm km/s}\) and \(\langle \lambda_C^{\parallel} \rangle/\langle \lambda_C^{\perp} \rangle = 1.18\) for \(V_{\rm SW} < 450~{\rm km/s}\).
    We report in Table \ref{tab:acewind4050} similar values to those of \citet{DassoEA2005} for slow wind but not for fast wind.
    A possible reason for the inconsistency in the fast wind ratio lies in the differences in methods of computing \(\lambda_C^{\parallel}\) and \(\lambda_C^{\perp}\).
    
    We provide individual radial power-law fits to results from Figure \ref{fig:corr_wide}; these are shown 
    in Table \ref{tab:rad_scaling} for different ranges of heliocentric distance.
    The more rapid radial evolution of \(\lambda_C^{\parallel}\) is reflected by the larger power-law exponents.
    Once \(\lambda_C^{\parallel}\) catches up to \(\lambda_C^{\perp}\), isotropy is roughly maintained, as can be interpreted by the combination of nearly equal power-law exponents for \(\lambda_C^{\parallel}\) and \(\lambda_C^{\perp}\) as well as their similar radial scaling for radial distances \(0.30~{\rm au} < R < 1~{\rm au}\).
  
    The connection between PSP and ACE/WIND and their general consistency presented in 
    Figures \ref{fig:corr_wide} and \ref{fig:ratio_wide} suggest a reversal of the initial anisotropy at radial distances within \(0.40~{\rm au}\).
    The combined ACE/WIND ratio \(\langle \lambda_C^{\parallel} \rangle / \langle \lambda_C^{\perp} \rangle > 1\) for all \(V_{\rm SW}\) conditions given in Table \ref{tab:acewind4050} confirms this observation.
    However, whether this reversal of the initial anisotropy is preserved past 1~au, or is just a temporary deviation from isotropy, remains inconclusive.
    One might presume from the radial scalings in Table \ref{tab:rad_scaling} for radial distances beyond \(1~{\rm au}\) that \(\lambda_C^{\parallel}\) continues to grow past the perpendicular scale.
    However, no strong conclusion can be drawn from the available data since very few parallel intervals are found in the Voyager data. Pickup ions are also expected to affect \(\lambda_C^\parallel\) above 5 au \citep{Zank2017ApJ}.

\section{Discussion}\label{sec:disc}

    Data from the first eight PSP encounters reveals an anisotropy with \(\lambda_C^{\parallel} < \lambda_C^{\perp}\) at \(\sim .08~{\rm au}\), also observed by \citet{BandyopadhyayMcComas2021}.
    This is likely explained by the physical size of granulated cells on the Sun's surface.
    The parallel scale is more dependent on the nature of the mechanisms that inject magnetic energy along the radial component of the magnetic field in the corona.
    In this regard it is possible to develop arguments \citep{MatthaeusEA1990JGR,ZankEA2021PoP} of a general nature that the scale of energy injection relative to the magnetic field exerts a strong influence on the corresponding correlation scales. 
    This reasoning may well explain the anisotropy observed in the inner heliosphere but there are no observations or detailed theories as yet that firmly establish this connection. 
    
    As one moves outward to \(\sim 0.40~{\rm au}\), \(\lambda_C^{\parallel}\) and \(\lambda_C^{\perp}\) isotropize.
    A deviation from isotropy occurs once the solar wind reaches \(1~{\rm au}\), as observed by PSP, ACE, and WIND.
    Near \(1~{\rm au}\), a reversal of the initial anisotropy is observed, such that the parallel correlation scale becomes the larger of the two, with \(\lambda_C^{\parallel}/\lambda_C^{\perp} = 1.28\) (see Table \ref{tab:acewind4050}).
    This is consistent with the observations by \citet{DassoEA2005}.

    The trend towards greater \(\lambda_C^{\parallel}/\lambda_C^{\perp}\) at \(1~{\rm au}\) and beyond might be interpreted in at least two different ways:
    
    (1) The anisotropy at \(1~{\rm au}\) represents a temporary deviation from the isotropy achieved at \(\approx 0.40~{\rm au}\).
    As a result, for heliocentric distances beyond \(1~{\rm au}\), \(\lambda_C^{\parallel}/\lambda_C^{\perp}\) remains \(\approx 1\), 
    even as other transient deviations occur, as seen
    in Figure \ref{fig:corr_wide}. 
    We note that turbulent MHD simulations support this view, finding that, after a startup transient which lasts several nonlinear times, the MHD system
    settles into a regime in which the correlation scale ratio remains roughly constant with values not far from unity \citep{Bandyopadhyay2019JFM}.
    Analysis of turbulence ``aging'' in the solar wind 
    \citep{Matthaeus1998JGR_turbulence_age} indicates that this condition should be well fulfilled within 1~au and beyond \citep[see also][]{ChhiberEA2016}. 
    
    (2) The anisotropy at \(1~{\rm au}\) marks a change in system dynamics that causes a more rapid increasing in \(\lambda_C^{\parallel}\) relative to \(\lambda_C^{\perp}\) beyond \(1~{\rm au}\).
    This can be supported by radial power-law fits yielding a stronger radial dependence in \(\lambda_C^{\parallel}\).
    A caveat for both these possibilities is the weak statistical weight of parallel intervals from Voyager 1.
    Therefore, we cannot draw any strong conclusions from the evolution of \(\lambda_C^{\parallel}/\lambda_C^{\perp}\) beyond \(1~{\rm au}\).
    
    A stronger basis for conclusions may emerge from analysis of the more populated data intervals in the range between  \(0.80~{\rm au} < R < 1~{\rm au}\) where one observes anisotropy increasing with radial distance.
    For example, there is ample evidence that solar wind turbulence has not yet attained a fully-developed character in the inner-heliosphere where there is evidence of increasing small-scale intermittency with increasing heliocentric distance \citep{AlbertiEA2020,TelloniEA2021,CuestaEA2022,SioulasEA2022}.
    However, near \(1~{\rm au}\), the rate of increase of intermittency reverses and decreases moving towards larger radial distances \citep{ParasharEA2019,CuestaEA2022}.
    If the solar wind is still evolving towards fully developed status near 1 au, then the outer-scale anisotropy -- what we have characterized by measuring 
    \(\lambda_C^{\parallel}/\lambda_C^{\perp}\) beyond \(1~{\rm au}\), may also still be evolving. In this sense the trend just inside of 1~au may represent the relatively slower evolution towards a weakly quasi-two dimensional state.
    Such anisotropy may be the consequence of enhanced formation of perpendicular gradients relative to the large-scale magnetic-field direction \citep{ShebalinEA1983}, even if that familiar anisotropy is more often associated with inertial range scales where the effect is of greater magnitude than the moderate departure from isotropy observed here in the outer heliosphere beyond 1 au. 
    
    Future PSP orbits will provide the opportunity to examine the evolution of turbulence correlations closer to the Sun's surface.
    The magnetic field direction and the solar wind direction are expected to be principally radial at the lower altitudes, so most observations will be of the parallel type, when standard Taylor hypothesis is applicable. 
    However lower solar wind speed, higher Alfv\'en speed and rapid spacecraft motion across the radial direction near perihelion may permit valuable studies of correlation anisotropy be carried out using modified forms of the Taylor hypothesis \citep{Matthaeus97,KleinEA15,PerezEA21}. 
    Large tangential velocities in the bulk flow close to the Sun \citep{Weber1967ApJ,Kasper2019Nature} may also permit evaluation of perpendicular correlations below the Alfv\'en transition region \citep{chhiber2022MNRAS}.

\section*{Acknowledgements}
    This research partially supported by NASA under the Heliospheric Supporting Research program grants
    80NSSC18K1210 and 80NSSC18K1648, by the Parker Solar Probe Guest Investigator program 80NSSC21K1765 at Delaware and 80NSSC21K1767 at Princeton, and by Heliophysics Guest Investigator program
    80NSSC19K0284, the 
    PSP/IS\(\odot\)IS\ Theory and Modeling project (Princeton 
    subcontract SUB0000165), and the PUNCH project under subcontract 
    NASA/SWRI N99054DS.

\appendix

\section{Instrumentation and Data Description}\label{sec:data}

    \textit{PSP.}
    Level 2 magnetic-field and Level 3 plasma data were extracted from NASA Goddard Space Flight Center Space Physics Data Facility (SPDF) in heliocentric RTN coordinates at full cadence.
    We use measurements made by the fluxgate magnetometer onboard the FIELDS instrument suite \citep{BaleEA2016} and by the Solar Probe Cup (SPC) onboard the Solar Wind Electrons Alphas and Protons (SWEAP) instrument \citep{KasperEA2016}.
    We use available data from the first 8 orbits, resampled as needed to the desired resolution of \({\rm 1s}\). 
    These data cover the time period between October 5, 2018 to June 30, 2021.
    Interval sizes vary by heliocentric distance, such that \(1~{\rm hr}\) and \(3~{\rm hr}\) intervals are used for \(R<0.30~{\rm au}\) and \(R>0.30~{\rm au}\), respectively (see discussion below). 
    A Hampel filter is applied to proton velocity components in real space to remove large outliers.
    For a definition of this Hampel filter, see \citet{Pearson2002}
    and \citet{BandyopadhyayEA2018}.
    When the solar wind speed in any PSP interval was unavailable due to data quality issues, speeds were linearly interpolated using nearby intervals in order to maximize the number of intervals for which 
    a correlation time can be converted to 
    length via the Taylor hypothesis.
    
    \textit{ACE.}
    Magnetic-field data was extracted from SPDF in RTN coordinates at a \(1~{\rm s}\) resolution for all available times from 5 February 1998 to 30 March 2008. 
    The 1~s resolution data were resampled via an averaging technique to a \(1~{\rm minute}\) resolution.
    The twin triaxial fluxgate magnetometer onboard ACE \citep{SmithEA1998} provides magnetic field data and the Solar Wind Election Proton Alpha Monitor instrument suite \citep{McComasEA1998} provides plasma data.
    An interval size of 1 day is utilized.
    From a total of 3707 intervals, 3576 intervals were useful.
    Here, a useful interval is defined to have no more than 80\% missing data.
    
    \textit{WIND.}
    Magnetic-field data were extracted from SPDF in RTN coordinates at a \(1~{\rm minute}\) resolution for all available times from 5 February 1998 to 5 February 2008.
    The Solar Wind Experiment \citep{OgilvieEA1995} provides plasma data and the Magnetic Field Investigation instrument suite \citep{LeppingEA1995} provides magnetic field data via a boom-mounted dual triaxial fluxgate magnetometer.
    An interval size of 1 day is utilized.
    From a total of 3650 intervals, 3626 intervals were useful.
    
    \textit{Voyager 1.}
    The existing archival Voyager 1 magnetic field data were extracted from SPDF in RTN coordinates at a \({\rm 1.92s}\) resolution.
    Data covers heliocentric distances ranging from 1 to 10~au.
    The Voyager mission used a dual low-field and high-field magnetometer system \citep{BehannonEA1977} for magnetic-field measurements, and the onboard plasma instrument \citep{BridgeEA1977} for other plasma parameters such as bulk flow speed and proton density.
    Although missing data and data gaps were encountered, we discovered several other inaccuracies with respect to the quality of the data, all of which were resolved \citep{MyMSThesis,CuestaEA2022}.
    The improved dataset is publicly available at \href{https://doi.org/10.5281/zenodo.5711177}{https://doi.org/10.5281/zenodo.5711177}.
    An interval size of 1 day is applied to all heliocentric distances covered by Voyager 1.
    From a total of 1333 intervals, 780 intervals were useful.
    
    \textit{Computation Specifics.}
    Helios 1 results presented in this study were extracted from \citet{RuizEA2011}.
    Results will not reflect selections of fast and slow solar wind, with the exception of reported statistics of combined ACE and WIND intervals given in Table \ref{tab:acewind4050}.
    We obtained \(\lambda_C^{\parallel}\) and \(\lambda_C^{\perp}\) from intervals with alignment angles assigned to the angular channels of \(0^{\circ}< \Theta < 40^{\circ}\) (\textit{parallel channel}) and \(50^{\circ}< \Theta <90^{\circ}\) (\textit{perpendicular channel}), respectively.
    When narrowing the angular ranges for parallel and perpendicular classifications by 10 degrees, the results remain nearly the same. 
    Further narrowing of the angular ranges produces large statistical uncertainties.
    
    Finally, we discuss the interval sizes used in this study. 
    For turbulence analyses, an interval size containing several correlation scales is ideal.
    Further, the interval should not be so large that solar rotation effects are included. 
    From \citet{IsaacsEA2015}, the optimized averaging window for computing correlation scales was determined to be between 10 and 20 hours long at \(1~{\rm au}\).
    Therefore, we decided that 1~day interval lengths were most suitable when computing correlation lengths for ACE, WIND, and Voyager 1.
    For PSP, we selected a boundary at \(0.30~{\rm au}\) to correspond to a shift from 1-hour to 3-hour intervals with increasing heliocentric distance. 
    The smaller intervals reflect the decreasing correlation scales with decrease in heliocentric distance \citep{ChhiberEA2021Dec}; the average correlation time at about \(0.10~{\rm au}\) is \(\approx 6~{\rm minutes}\), which allows for the minimum required oversampling \citep{IsaacsEA2015}.
    Additionally, closer to the sun, the local magnetic field is dominantly radial resulting in less opportunities to find longer intervals having perpendicular alignment angles.
    In order to obtain more statistical weight of perpendicular intervals at PSP perihelia, 1-hour interval times were selected \citep[see also][]{BandyopadhyayEA2022Feb}.




\newpage
\newpage
\newpage
\newpage

\bibliographystyle{aasjournal}




\end{document}